\documentclass[twocolumn,showpacs,aps,prl,superscriptaddress]{revtex4}
\usepackage{graphicx}
\usepackage{lineno}
\usepackage{dcolumn}
\usepackage{amsmath}
\usepackage{epsfig}

\RequirePackage{xspace}

\usepackage{relsize}
\def\babar{\mbox{\slshape B\kern-0.1em{\smaller A}\kern-0.1em
    B\kern-0.1em{\smaller A\kern-0.2em R}}}

\def\epem       {\ensuremath{e^+e^-}\xspace}

\def\piz   {\ensuremath{\pi^0}\xspace}

\def\pip   {\ensuremath{\pi^+}\xspace}
\def\pim   {\ensuremath{\pi^-}\xspace}

\def\pipm  {\ensuremath{\pi^\pm}\xspace}
\def\pimp  {\ensuremath{\pi^\mp}\xspace}

\def\Kbar  {\kern 0.2em\overline{\kern -0.2em K}{}\xspace}

\def\Kz    {\ensuremath{K^0}\xspace}
\def\Kzb   {\ensuremath{\Kbar^0}\xspace}
\def\KzKzb {\ensuremath{\Kz \kern -0.16em \Kzb}\xspace}
\def\Kp    {\ensuremath{K^+}\xspace}
\def\Km    {\ensuremath{K^-}\xspace}
\def\Kpm   {\ensuremath{K^\pm}\xspace}
\def\Kmp   {\ensuremath{K^\mp}\xspace}
\def\KpKm  {\ensuremath{\Kp \kern -0.16em \Km}\xspace}

\def\Dbar    {\kern 0.2em\overline{\kern -0.2em D}{}\xspace}

\def\Dz      {\ensuremath{D^0}\xspace}
\def\Dzb     {\ensuremath{\Dbar^0}\xspace}
\def\DzDzb   {\ensuremath{\Dz {\kern -0.16em \Dzb}}\xspace}
\def\Dp      {\ensuremath{D^+}\xspace}
\def\Dm      {\ensuremath{D^-}\xspace}

\def\DpDm    {\ensuremath{\Dp {\kern -0.16em \Dm}}\xspace}

\def\B       {\ensuremath{B}\xspace}
\def\Bbar    {\kern 0.18em\overline{\kern -0.18em B}{}\xspace}

\def\BB      {\ensuremath{B\Bbar}\xspace} 
\def\Bz      {\ensuremath{B^0}\xspace}
\def\Bzb     {\ensuremath{\Bbar^0}\xspace}
\def\BzBzb   {\ensuremath{\Bz {\kern -0.16em \Bzb}}\xspace}
\def\Bu      {\ensuremath{B^+}\xspace}
\def\Bub     {\ensuremath{B^-}\xspace}

\def\BpBm    {\ensuremath{\Bu {\kern -0.16em \Bub}}\xspace}

\def\BorBbar    {\kern 0.18em\optbar{\kern -0.18em B}{}\xspace}
\def\DorDbar    {\kern 0.18em\optbar{\kern -0.18em D}{}\xspace}
\def\KorKbar    {\kern 0.18em\optbar{\kern -0.18em K}{}\xspace}

\mathchardef\Upsilon="7107
\def\Y#1S{\ensuremath{\Upsilon{(#1S)}}\xspace}

\mathchardef\Deltares="7101
\mathchardef\Xi="7104
\mathchardef\Lambda="7103
\mathchardef\Sigma="7106
\mathchardef\Omega="710A

\def\Deltabar{\kern 0.25em\overline{\kern -0.25em \Deltares}{}\xspace}
\def\Lbar{\kern 0.2em\overline{\kern -0.2em\Lambda\kern 0.05em}\kern-0.05em{}\xspace}
\def\Sigbar{\kern 0.2em\overline{\kern -0.2em \Sigma}{}\xspace}
\def\Xibar{\kern 0.2em\overline{\kern -0.2em \Xi}{}\xspace}
\def\Obar{\kern 0.2em\overline{\kern -0.2em \Omega}{}\xspace}
\def\Nbar{\kern 0.2em\overline{\kern -0.2em N}{}\xspace}
\def\Xb{\kern 0.2em\overline{\kern -0.2em X}{}\xspace}

\def\mes        {\mbox{$m_{\rm ES}$}\xspace}

\newcommand{\tev}{\ensuremath{\mathrm{\,Te\kern -0.1em V}}\xspace}
\newcommand{\gev}{\ensuremath{\mathrm{\,Ge\kern -0.1em V}}\xspace}
\newcommand{\mev}{\ensuremath{\mathrm{\,Me\kern -0.1em V}}\xspace}
\newcommand{\kev}{\ensuremath{\mathrm{\,ke\kern -0.1em V}}\xspace}
\newcommand{\ev}{\ensuremath{\mathrm{\,e\kern -0.1em V}}\xspace}
\newcommand{\gevc}{\ensuremath{{\mathrm{\,Ge\kern -0.1em V\!/}c}}\xspace}
\newcommand{\mevc}{\ensuremath{{\mathrm{\,Me\kern -0.1em V\!/}c}}\xspace}
\newcommand{\gevcc}{\ensuremath{{\mathrm{\,Ge\kern -0.1em V\!/}c^2}}\xspace}
\newcommand{\mevcc}{\ensuremath{{\mathrm{\,Me\kern -0.1em V\!/}c^2}}\xspace}

\def\mus  {\ensuremath{\rm \,\mus}\xspace}

\def\ps   {\ensuremath{\rm \,ps}\xspace}

\def\mus        {\ensuremath{\,\mu{\rm s}}\xspace}    
\def\ps         {\ensuremath{{\rm \,ps}}\xspace}  

\def\to                 {\ensuremath{\rightarrow}\xspace}

\def\pep2{PEP-II}

\def\gsim{{~\raise.15em\hbox{$>$}\kern-.85em
          \lower.35em\hbox{$\sim$}~}\xspace}
\def\lsim{{~\raise.15em\hbox{$<$}\kern-.85em
          \lower.35em\hbox{$\sim$}~}\xspace}

\def\CP                {\ensuremath{C\!P}\xspace}

\def\stwoa{\ensuremath{\sin\! 2 \alpha  }\xspace}

\def\deltat{\ensuremath{{\rm \Delta}t}\xspace}
\def\deltamd{\ensuremath{{\rm \Delta}m_d}\xspace}

\xspace

\newcommand{\epjBase}        {Eur.\ Phys.\ Jour.\xspace}
\newcommand{\jprlBase}       {Phys.\ Rev.\ Lett.\xspace}
\newcommand{\jprBase}        {Phys.\ Rev.\xspace}
\newcommand{\jplBase}        {Phys.\ Lett.\xspace}
\newcommand{\epjc}      [1]  {\epjBase\ C~{\bf #1}}

\newcommand{\plb}       [1]  {\jplBase\ B~{\bf #1}}

\newcommand{\jprl}      [1]  {\jprlBase\ {\bf #1}}
\newcommand{\jprd}      [1]  {\jprBase\ D~{\bf #1}}

\newcommand{\progtp}    [1]  {{Prog.\ Theor.\ Phys.\ {\bf #1}}}

\def\jetset74   {\mbox{\tt Jetset \hspace{-0.5em}7.\hspace{-0.2em}4}\xspace}

\def\fish    {\ensuremath{\cal F}}
\def\akpi  {\ensuremath{{\cal A}_{K\pi}}}

\def\fpm {\ensuremath{f_{\pm}(\deltat)}}

\def\spipi {\ensuremath{S_{\pi\pi}}}
\def\cpipi {\ensuremath{C_{\pi\pi}}}
\def\de {\ensuremath{\Delta E}}

\def\Btag {\ensuremath{B_{\rm tag}}}

\def\Bflav {\ensuremath{B_{\rm flav}}}

\RequirePackage{xspace}

\def\figurebox#1#2#3{%
    \def\arg{#3}%
    \ifx\arg\empty
    {\hfill\vbox{\hsize#2\hrule\hbox to #2{\vrule\hfill\vbox to #1{\hsize#2\vfill}\vrule}\hrule}\hfill}%
    \else
    {\hfill\epsfbox{#3}\hfill}%
    \fi}

\long\def\inst#1{\par\nobreak\kern 4pt\nobreak
    {\it #1}\par\vskip 10pt plus 3pt minus 3pt}

\begin{document}

\title{
{
\Large \bf \boldmath
Improved Measurements of \CP-Violating\\
 Asymmetry Amplitudes in $\Bz\to\pip\pim$ Decays
}}

%
\author{B.~Aubert}
\author{R.~Barate}
\author{D.~Boutigny}
\author{F.~Couderc}
\author{Y.~Karyotakis}
\author{J.~P.~Lees}
\author{V.~Poireau}
\author{V.~Tisserand}
\author{A.~Zghiche}
\affiliation{Laboratoire de Physique des Particules, F-74941 Annecy-le-Vieux, France }
\author{E.~Grauges-Pous}
\affiliation{IFAE, Universitat Autonoma de Barcelona, E-08193 Bellaterra, Barcelona, Spain }
% ...
\collaboration{The \babar\ Collaboration}
\noaffiliation

\begin{abstract}
We present updated measurements of the \CP-violating parameters 
$\spipi$ and $\cpipi$ in $\Bz\to\pip\pim$ decays. Using a sample of 
$227$ million $\Y4S\to\BB$ decays collected with the \babar\ detector 
at the \pep2\ asymmetric-energy $\epem$ collider at SLAC, we observe 
$467\pm 33$ signal decays and measure 
$\spipi = -0.30\pm 0.17\,({\rm stat})\pm 0.03\,({\rm syst})$, and 
$\cpipi = -0.09\pm 0.15\,({\rm stat})\pm 0.04\,({\rm syst})$.
\end{abstract}


\maketitle

In the standard model, \CP-violating effects in the $B$-meson system arise from a
 single phase in the Cabibbo-Kobayashi-Maskawa quark-mixing matrix~\cite{CKM}.  
 In this context, neutral-$B$ decays to the \CP\ eigenstate 
$\pip\pim$ can exhibit mixing-induced \CP\ violation through interference between 
decays with and without $\Bz$--$\Bzb$ mixing, and direct \CP\ violation through 
interference between the $b\to u$ tree and $b\to d$ penguin decay processes~\cite{pipicpv}.  
Both effects are observable in the time evolution of the asymmetry between $\Bz$ and $\Bzb$ 
decays to $\pip\pim$, where mixing-induced \CP\ violation leads to a sine term 
with amplitude $\spipi$ and direct \CP\ violation leads to a cosine term with 
amplitude $\cpipi$.  In the absence of the penguin process, $\cpipi = 0$ and $\spipi = \stwoa$, with
$\alpha \equiv \arg\left[-V_{\rm td}^{}V_{\rm tb}^{*}/V_{\rm ud}^{}V_{\rm ub}^{*}\right]$,
while significant tree-penguin interference leads to $\spipi = \sqrt{1 - \cpipi^2}\sin{2\alpha_{\rm eff}}$, 
where $\alpha_{\rm eff}$ is the effective value of $\alpha$ and $\cpipi\ne 0$ if the 
strong phases of the tree and penguin decay amplitudes are different.
The difference $\Delta\alpha_{\pi\pi}\equiv \alpha - \alpha_{\rm eff}$ can be 
determined from a model-independent analysis using the isospin-related decays 
$B^{\pm}\to\pipm\piz$ and 
$\Bz,\,\Bzb\to\piz\piz$~\cite{alphafrompenguins,isospin}.

The Belle collaboration recently reported~\cite{BelleSin2alpha2004} 
an observation of \CP\ violation in $\Bz\to\pip\pim$ decays using a data sample
of $152$ million $\BB$ pairs, while our previous measurement~\cite{BaBarSin2alpha2002} 
on a sample of $88$ million $\BB$ pairs was consistent with no \CP\ violation.  In this 
paper we report improved measurements of the \CP-violating parameters $\spipi$ and 
$\cpipi$ using a data sample comprising $227$ million $\BB$ pairs collected with the 
\babar\ detector at the \pep2\ asymmetric-energy $\epem$ collider at SLAC.

The \babar\ detector is described in detail elsewhere~\cite{ref:babar}.  
The primary components used in this analysis are a 
charged-particle tracking system consisting of a five-layer silicon 
vertex tracker (SVT) and a 40-layer drift chamber (DCH) surrounded 
by a $1.5$-T solenoidal magnet, an electromagnetic calorimeter 
(EMC) comprising $6580$ CsI(Tl) crystals, and a detector of 
internally reflected Cherenkov light (DIRC) providing $K$--$\pi$ 
separation over the range of laboratory momentum relevant
for this analysis ($1.5$--$4.5\gevc$).

The analysis method is similar to that used in our previous measurement of 
$\spipi$ and $\cpipi$~\cite{BaBarSin2alpha2002}.  We reconstruct a sample of 
neutral $B$ mesons ($B_{\rm rec}$) decaying to final states with two charged 
tracks, and examine the remaining particles in each event to infer whether 
the second $B$ meson (\Btag) decayed as a $\Bz$ or $\Bzb$ (flavor tag).
We first perform a maximum-likelihood fit that uses kinematic, event-shape, 
and particle-identification information 
to determine signal and background yields corresponding to the four 
distinguishable final states ($\pip\pim$, $\Kp\pim$, $\Km\pip$, $\Kp\Km$).  
The results of
this fit are described in Ref.~\cite{BaBarAkpiPRL}, which reports the first evidence of
direct \CP\ violation in $\Bz\to\Kp\pim$ decays~\cite{BelleKpiRef}.  The 
\CP\ asymmetry parameters
in $\Bz\to\pip\pim$ decays are then determined from a second fit including
information about the flavor of $B_{\rm tag}$ and the difference $\deltat$
between the decay times of the $B_{\rm rec}$ and $B_{\rm tag}$ decays.  
The decay rate distribution $f_+\,(f_-)$ when $B_{\rm rec}\to\pip\pim$ and 
$\Btag = \Bz\,(\Bzb)$ is given by
\begin{eqnarray}
\fpm = \frac{e^{-\left|\deltat\right|/\tau}}{4\tau} [1
& \pm & \spipi\sin(\deltamd\deltat) \nonumber \\
& \mp & \cpipi\cos(\deltamd\deltat)],
\label{fplusminus}
\end{eqnarray}
where $\tau$ is the $\Bz$ lifetime and $\deltamd$ is the mixing
frequency due to the neutral-$B$-meson eigenstate mass difference.

The analysis begins by reconstructing two-body neutral-$B$ decays from pairs of 
oppositely-charged tracks found within the geometric acceptance of the DIRC
and originating from a common decay point near the interaction region.
We require that each track have an associated Cherenkov angle ($\theta_c$) 
measured with at least five signal photons detected in the DIRC; the 
value of $\theta_c$ must agree within four standard deviations ($\sigma$) with 
either the pion or kaon particle hypothesis.  The last requirement efficiently 
removes events with high-momentum protons.  Electrons are removed based on 
energy-loss measurements in the SVT and DCH, and on a comparison of the track momentum 
and associated energy deposited in the EMC.  

Identification of pions and kaons is primarily accomplished by including 
$\theta_c$ as a discriminating variable in the 
maximum likelihood fit.  We construct probability density functions (PDFs) 
for $\theta_c$ from a sample of approximately 
$430000$ $D^{*+}\to D^0\pi^+\,(\Dz\to\Km\pip)$ decays reconstructed in data, where 
$\Kmp/\pipm$ tracks are identified through the charge correlation with the
$\pipm$ from the $D^{*\pm}$ decay.  Although we find no systematic difference between
positive and negative tracks, the PDFs are constructed separately for $\Kp$, $\Km$, 
$\pip$, and $\pim$ tracks as a function of momentum and 
polar angle using the measured and expected values of $\theta_c$, and the uncertainty.

Signal decays are identified using two kinematic variables: (1) the 
difference $\de$ between the reconstructed energy of the $B$ candidate
in the $\epem$ center-of-mass (CM) frame and $\sqrt{s}/2$, and (2) the beam-energy 
substituted mass 
$\mes = \sqrt{(s/2 + {\mathbf {p}}_i\cdot {\mathbf {p}}_B)^2/E_i^2- {\mathbf {p}}_B^2}$.
Here, $\sqrt{s}$ is the total CM energy, and the $B$ momentum ${\mathbf {p_B}}$ 
and the four-momentum $(E_i, {\mathbf {p_i}})$ of the $\epem$ initial state are 
defined in the laboratory frame. 
For signal decays, $\de$ and $\mes$ have Gaussian
distributions with standard deviations of $27\mev$ and $2.6\mevcc$, respectively.  
The distribution of $\mes$ peaks near the $B$ mass for all four final states.  
To simplify the likelihood fit, we reconstruct the kinematics 
of the $B$ candidate using the pion mass for all tracks.  With this choice, 
$\Bz\to\pip\pim$ decays peak near $\de = 0$.  For $B$ decays with one or two 
kaons in the final state, the $\de$ peak position is shifted and parameterized 
as a function of the kaon momentum in the laboratory frame.  The average shifts 
with respect to zero are $-45\mev$ and $-91\mev$, respectively, and this 
separation in $\de$ provides additional discriminating power in the fit.  
We require $5.20 < \mes < 5.29\gevcc$ and 
$\left|\de\right|<150\mev$.  The large sideband region in $\mes$ is used to 
determine background-shape parameters, while the wide range in $\de$ allows us 
to separate $B$ decays to all four final states in the same fit.

We have studied potential backgrounds from higher-multiplicity $B$ decays 
and find them to be negligible near $\de = 0$.  The dominant source of background 
is the process $\epem\to q\bar{q}\; (q=u,d,s,c)$, which produces a distinctive 
jet-like topology.  In the CM frame we define the angle $\theta_S$ between the 
sphericity axis~\cite{sph} of the $B$ candidate and the sphericity axis of the 
remaining particles in the event.  For background events, $\left|\cos{\theta_S}\right|$ 
peaks sharply near unity, while it is nearly flat for signal decays.  We require 
$\left|\cos{\theta_S}\right|<0.8$, which removes approximately $80\%$ of this 
background.  Additional background suppression is accomplished by
a Fisher discriminant ${\cal F}$~\cite{BaBarSin2alpha2002} based on
the momentum flow relative to the $\pip\pim$ thrust axis of all tracks and clusters
in the event, excluding the $\pi\pi$ pair.  We use ${\cal F}$ as an additional 
discriminating variable in the fit.

We use a multivariate technique~\cite{BaBarsin2beta} to determine the flavor of 
the $\Btag$ meson.  Separate neural networks are trained to identify primary leptons, 
kaons, soft pions from $D^*$ decays, and high-momentum charged particles from \B\ decays.  
Events are assigned to one of five mutually exclusive tagging categories 
based on the estimated average mistag probability and the source of the tagging information.
The quality of tagging is expressed in terms of the effective efficiency 
$Q = \sum_k \epsilon_k (1-2w_k)^2$, where $\epsilon_k$ and $w_k$ are the 
efficiencies and mistag probabilities for events tagged in category $k$.
We measure the tagging performance in a data sample \Bflav\ of fully reconstructed neutral 
$B$ decays to $D^{(*)-}(\pip,\, \rho^+,\, a_1^+)$, and find a total effective 
efficiency of $Q = 29.9\pm 0.5$.  The assumption of equal tagging efficiencies and 
mistag probabilities for signal $\pip\pim$, $\Kp\pim$, and $\Kp\Km$ decays is validated 
in a detailed Monte Carlo simulation.  Separate background efficiencies for the different 
decay modes are determined simultaneously with $\spipi$ and $\cpipi$ in the fit.

The time difference $\deltat \equiv \Delta z/\beta\gamma c$ is obtained from the 
known boost of the $\epem$ system ($\beta\gamma = 0.56$) and the measured distance 
$\Delta z$ along the beam ($z$) axis between the $B_{\rm rec}$ and $B_{\rm tag}$ decay
vertices.  We require $\left|\deltat\right|<20\ps$ and 
$\sigma_{\deltat} < 2.5\ps$, where $\sigma_{\deltat}$ is the uncertainty on $\deltat$ determined
separately for each event.  The resolution function for signal candidates is a sum of 
three Gaussians, identical to the one described in Ref.~\cite{BaBarsin2beta}, 
with parameters determined from a fit to the \Bflav\ sample (including events in all 
five tagging categories).  The background $\deltat$ distribution is modeled
as the sum of three Gaussian functions, where the common parameters used to describe the
background shape for all tagging categories are determined simultaneously with 
the \CP\ parameters in the maximum likelihood fit.

We use an unbinned extended maximum likelihood fit to extract \CP\ parameters
from the $B_{\rm rec}$ sample.  The likelihood for candidate $j$ tagged in category 
$k$ is obtained by summing the product of event yield $n_{i}$, tagging efficiency 
$\epsilon_{i,k}$, and probability ${\cal P}_{i,k}$ over the eight possible signal 
and background hypotheses $i$ (referring to $\pi^{+}\pi^{-}$, $K^{+}\pi^{-}$, 
$K^{-}\pi^{+}$, and $K^{+}K^{-}$ combinations).  The extended likelihood function 
for category $k$ is
\begin{equation}
{\cal L}_k = \exp{\left(-\sum_{i}n_i\epsilon_{i,k}\right)}
\prod_{j}\left[\sum_{i}n_i\epsilon_{i,k}{\cal P}_{i,k}(\vec{x}_j;\vec{\alpha}_i)\right].
\end{equation}
The yields for the $K\pi$ final state are parameterized as 
$n_{\Kpm\pimp}=n_{K\pi}\left(1\mp \akpi\right)/2$, where $\akpi$
is the direct-\CP-violating asymmetry~\cite{BaBarAkpiPRL}.
The probabilities ${\cal P}_{i,k}$ are evaluated as the product of PDFs 
for each of the independent variables 
$\vec{x}_j = \left\{\mes, \de, {\cal F}, \theta_c^+, \theta_c^-, \deltat\right\}$
with parameters $\vec{\alpha}_i$,
where $\theta_c^+$ and $\theta_c^-$ are the Cherenkov angles for the positively- and 
negatively-charged tracks.  The $\deltat$ PDF for signal $\pip\pim$ decays is given 
by Eq.~\ref{fplusminus} modified to include the mistag probabilities for each tag 
category, and convolved with the signal resolution function.  The $\deltat$ PDF for 
signal $K\pi$ decays takes into account $\Bz$--$\Bzb$ mixing and the correlation 
between the charge of the kaon and the flavor of $\B_{\rm tag}$.  We fix $\tau$ and 
$\deltamd$ to their world-average values~\cite{PDG2004}.  The total likelihood 
${\cal L}$ is the product of likelihoods for each tagging category, and the free 
parameters are determined by maximizing the quantity $\ln{\cal L}$.

The fit proceeds in two steps.  First, the signal and background yields and $K\pi$ 
charge asymmetries are determined in a separate fit that does not use flavor-tagging or 
$\deltat$ information~\cite{BaBarAkpiPRL}.  Out of a fitted sample of $68030$ events, 
we find $n_{\pi\pi} = 467\pm 33$, $n_{K\pi} = 1606\pm 51$, and $n_{KK}=3\pm 12$ decays, 
and measure $\akpi = -0.133\pm 0.030$, where all errors are statistical only.  We next
add the flavor tagging and $\deltat$ information and perform a fit for $\spipi$
and $\cpipi$.  We fix the signal and background yields and charge asymmetries
to values determined in the first fit, and fix the signal parameters describing
flavor-tagging and $\deltat$ resolution function parameters to the values determined
in the $B_{\rm flav}$ sample.  By fixing these parameters we reduce the total number
of free parameters by $30$ relative to our previous analysis~\cite{BaBarSin2alpha2002}.
A total of $46$ parameters are left free in the fit, including $12$ parameters describing 
the background PDFs for $\mes$, $\de$, and $\fish$; $8$ parameters describing the background 
$\deltat$ PDF; $12$ background flavor-tagging efficiencies; $12$ background flavor-tagging 
efficiency asymmetries; and $\spipi$ and $\cpipi$.  The fit yields
\begin{eqnarray*}
\spipi & =          & -0.30\pm 0.17\,({\rm stat})\pm 0.03\,({\rm syst}),\\
\cpipi & =          & -0.09\pm 0.15\,({\rm stat})\pm 0.04\,({\rm syst}),
\end{eqnarray*}
where the correlation between $\spipi$ and $\cpipi$ is $-1.6\%$, and the
correlations with all other free parameters are less than $1\%$.  These 
values are consistent with, and supersede, our previously published 
measurements~\cite{BaBarSin2alpha2002}.

\begin{figure}[!tbp]
\begin{center}
\includegraphics[width=0.45\linewidth]{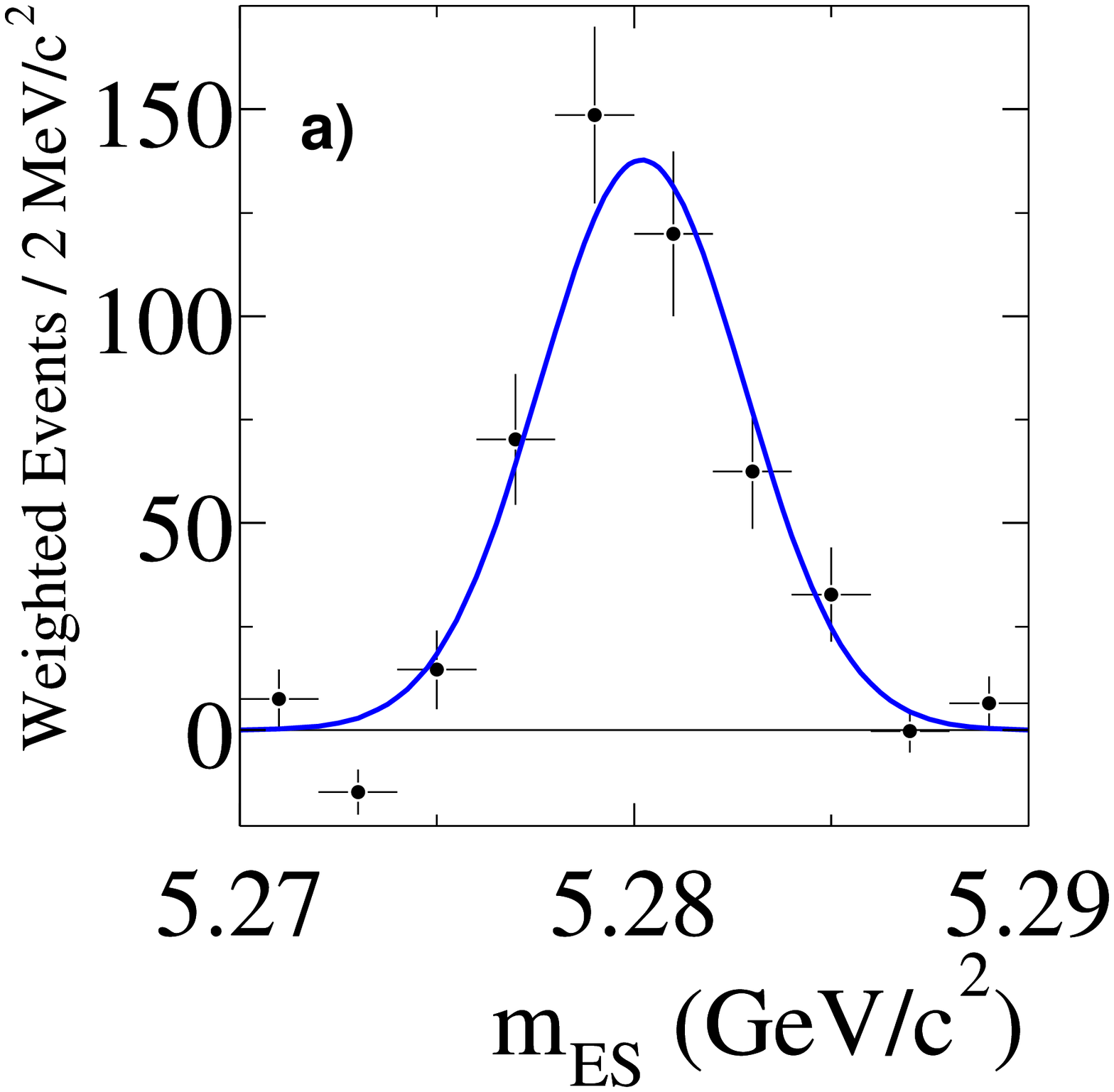}\hfill\includegraphics[width=0.45\linewidth]{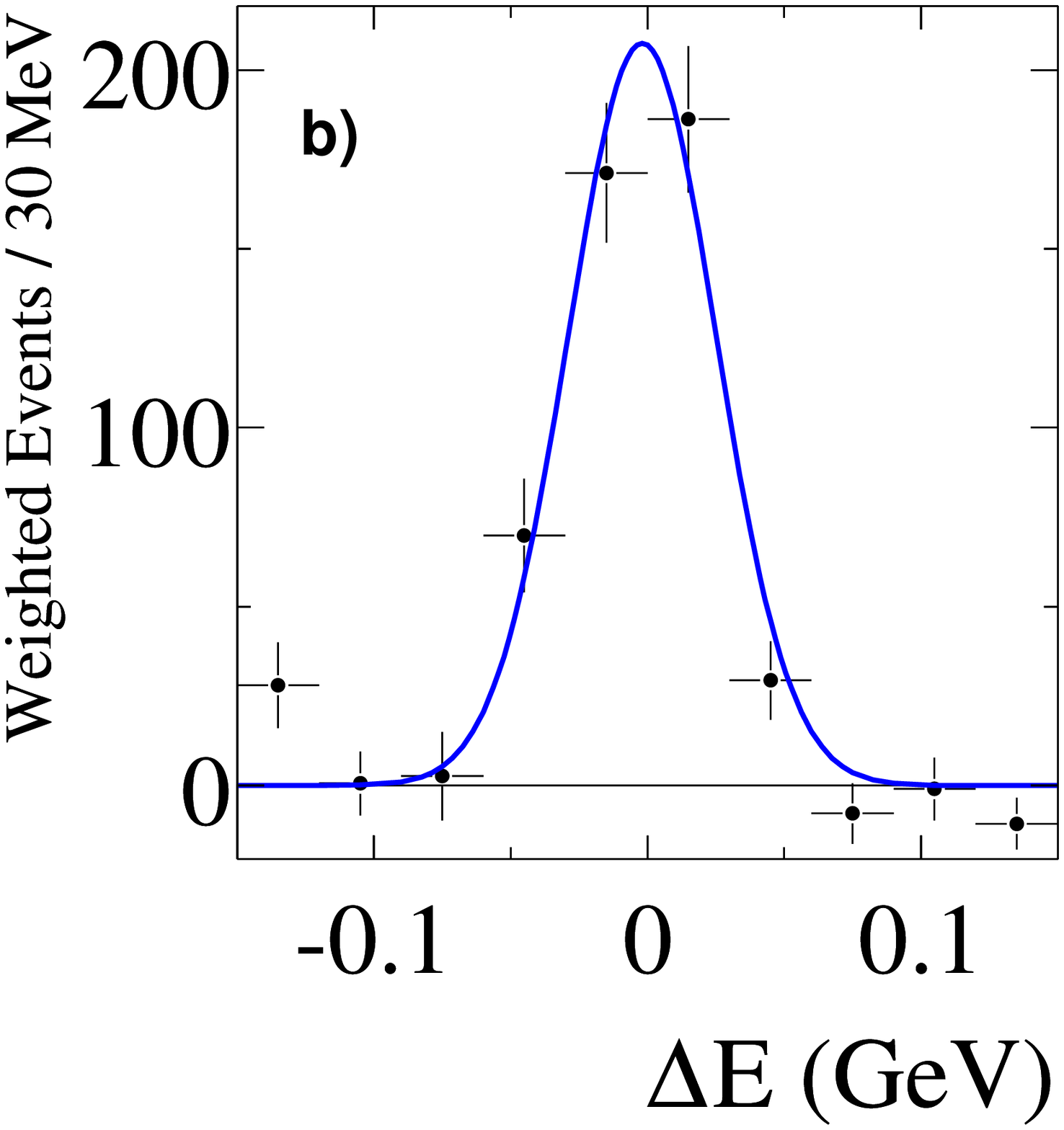}
\vskip1mm
\includegraphics[width=0.45\linewidth]{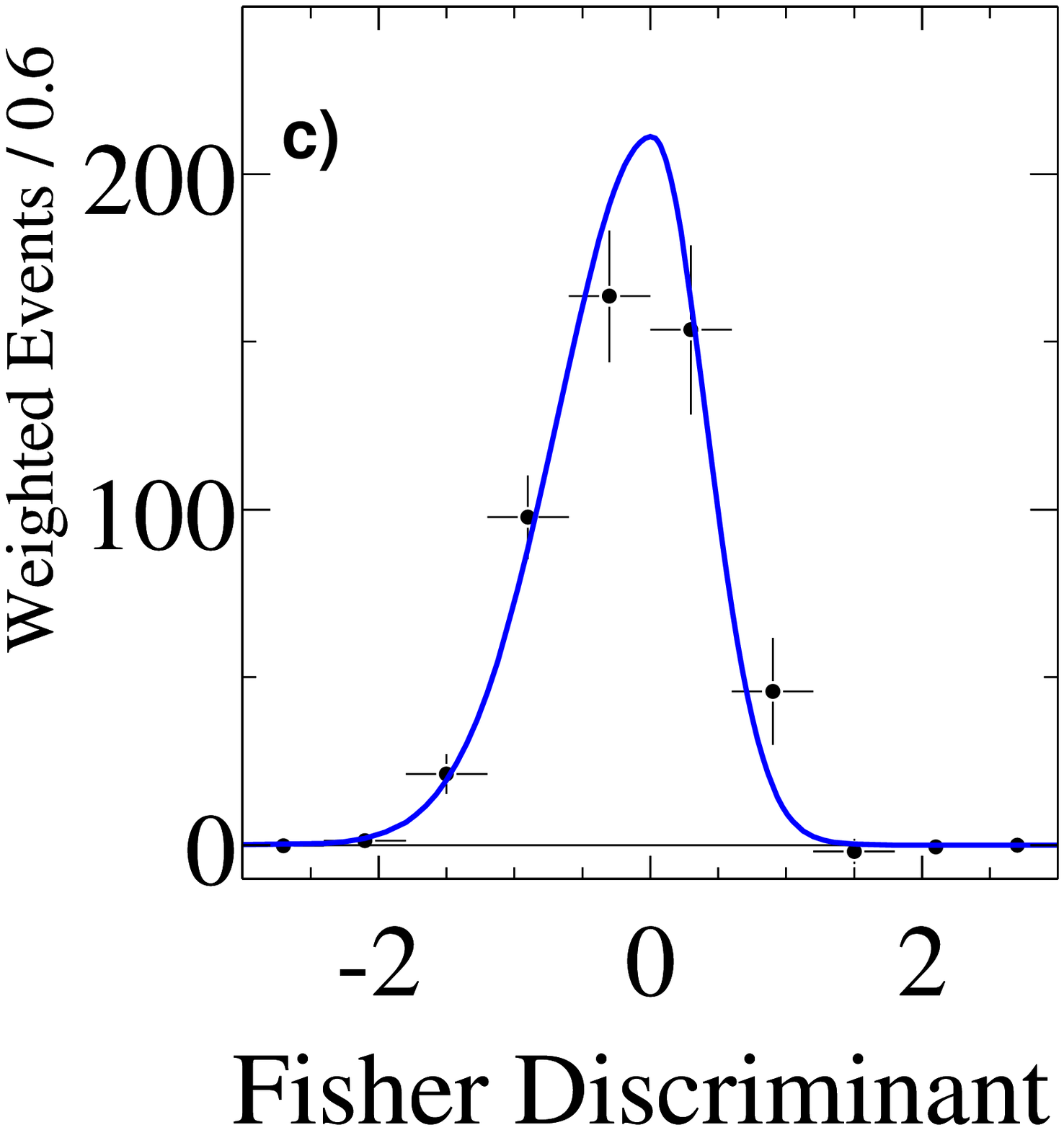}\hfill\includegraphics[width=0.45\linewidth]{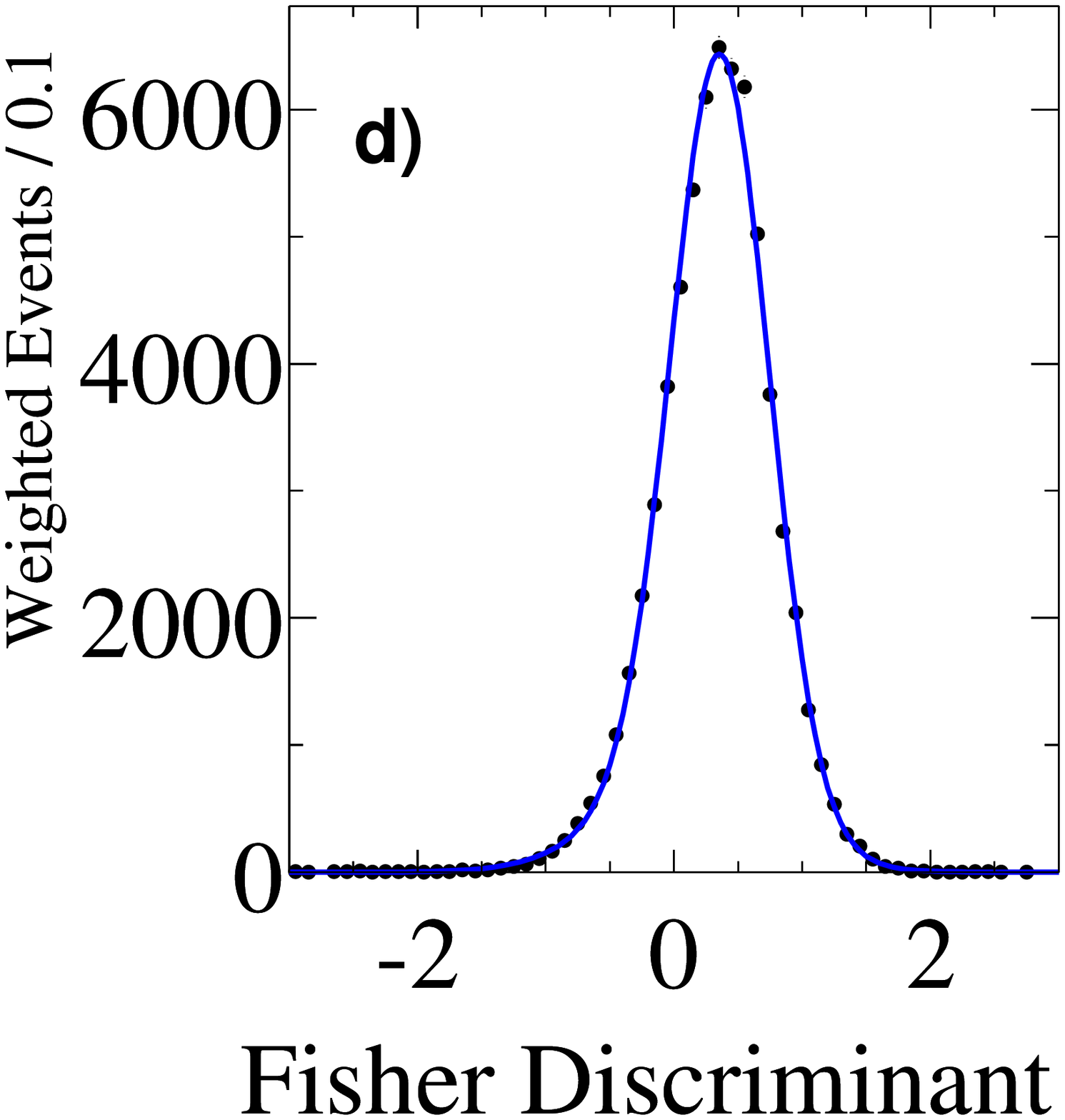}
\caption{Distributions of (a) $\mes$, (b) $\de$, and (c) ${\cal F}$ 
for signal $\pip\pim$ events (points with error bars), and (d) the distribution
of ${\cal F}$ for $q\bar{q}$ background events, using the weighting technique 
described in Ref.~\cite{sPlots}.
Solid curves represent the corresponding PDFs used in the fit.}
\label{fig:pipi}
\end{center}
\end{figure}

\begin{figure}[!tbp]
\begin{center}
\includegraphics[width=0.55\linewidth]{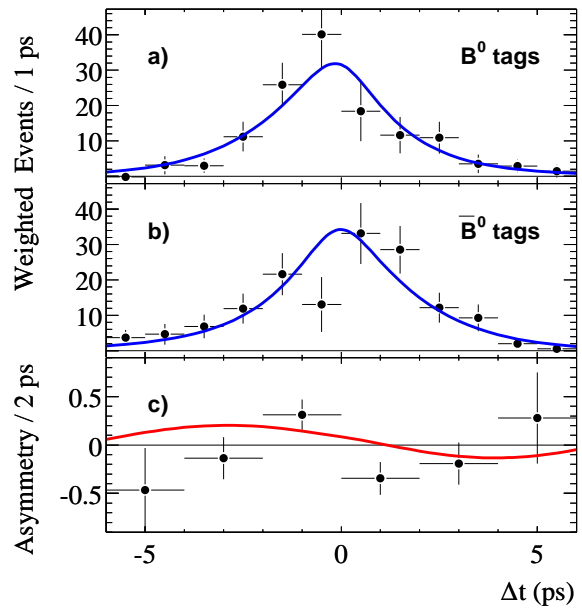}
\caption{Distributions of the decay-time difference $\deltat$ using
the event-weighting technique described in the text.
The top two plots show events where $B_{\rm tag}$ is
identified as (a) $\Bz$ ($n_{\Bz}$) or (b) $\Bzb$ ($n_{\Bzb}$), where
the solid curves indicate the signal PDFs used in the fit. 
(c) The asymmetry (points with errors), defined as 
$\left(n_{\Bz} - n_{\Bzb}\right)/\left(n_{\Bz} + n_{\Bzb}\right)$, for signal
events in each $\deltat$ bin, and the projection of the fit (solid curve).}
\label{fig:asym}
\end{center}
\end{figure}

We use the event-weighting technique described in Ref.~\cite{sPlots} to check
the agreement between PDFs and data for signal $\pip\pim$ candidates.  For Figs.~\ref{fig:pipi}(a-c), we perform a fit excluding the variable being plotted, and the covariance matrix is used to determine a weight (probability) that each event is signal, not background.  The resulting distributions 
(points with errors) are normalized to the signal yield ($467$) and can be directly compared with the 
PDFs (solid curves) used in the fit for $\spipi$ and $\cpipi$.  In Fig.~\ref{fig:pipi}d, we use a similar technique to compare the ${\cal F}$ distribution based on the probability to be a $q\bar{q}$ event with the PDF used for background events.
Using the same event-weighting technique, in Fig.~\ref{fig:asym} we show distributions of $\deltat$ for signal 
$\pip\pim$ events with $B_{\rm tag}$ tagged as $\Bz$ or $\Bzb$, and the asymmetry as a 
function of $\deltat$.  In all cases, we find good agreement between data and PDFs.

As a consistency check on the $\deltat$ resolution function, we 
take advantage of the large number of $K\pi$ signal decays in the $B_{\rm rec}$ sample
to perform a $\Bz$--$\Bzb$ mixing analysis.  Floating $\tau$ and $\deltamd$ along 
with $\spipi$, $\cpipi$, and $\akpi$, we find values consistent with the world averages 
($\tau = 1.60\pm 0.04\,{\rm ps}$ and $\deltamd = 0.523\pm 0.028\,{\rm ps}^{-1}$), 
and \CP parameters consistent with the nominal fit results.  This test 
gives us confidence that the $\deltat$ measurement is unbiased.

\begin{figure}[!tbp]
\begin{center}
\includegraphics[width=0.8\linewidth]{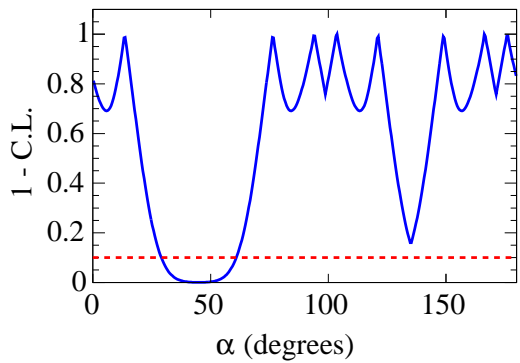}
\caption{Constraints on $\alpha$ derived from the isospin analysis using 
$\spipi$, $\cpipi$, and $\Delta\alpha_{\pi\pi}$ (Ref.~\cite{isospin}).  Values of $\alpha$
for which the solid line lies below the dashed line are excluded at $90\%$ C.L.}
\label{fig:alpha}
\end{center}
\end{figure}

The dominant sources of systematic uncertainty include imperfect knowledge of the PDF shape 
parameters; the $B$-flavor-tagging parameters; the alignment
of the SVT; the event-by-event beam-spot position; 
the potential effect of doubly Cabibbo-suppressed decays of the $B_{\rm tag}$ 
meson~\cite{Owen}, and the $B$ lifetime and mixing frequency.  
We verify that we are sensitive to non-zero values of $\spipi$ and $\cpipi$ by fitting 
a large sample of Monte-Carlo simulated signal decays with large values of the 
\CP\ parameters.  Although the fit results are consistent with the generated values, 
we assign the sum in quadrature of the statistical uncertainty and the difference between 
the fitted and generated values as a conservative 
systematic error accounting for potential bias in the fit procedure.  The effect of 
uncertainty in the signal and background yields and $K\pi$ asymmetries is negligible for 
both $\spipi$ and $\cpipi$.  The total systematic uncertainty is calculated by summing in 
quadrature the individual contributions.

Using the model-independent isospin 
analysis~\cite{alphafrompenguins} (neglecting electroweak penguin 
amplitudes) and the technique described in Ref.~\cite{ckmfitter}, we 
display in Fig.~\ref{fig:alpha} the confidence level (C.L.) derived from the
measured values of $\spipi$ and $\cpipi$ reported here, and 
the results for $\Delta\alpha_{\pi\pi}$ determined in Ref.~\cite{isospin}.
Values of $\alpha$ in the range $\left[29^{\circ},61^{\circ} \right]$ are excluded at the 
$90\%$ C.L.

In summary, we present improved measurements of the \CP-violating 
asymmetry amplitudes $\spipi$ and $\cpipi$, which govern the time distributions of 
$\Bz\to\pip\pim$ decays.  We find $\spipi = -0.30\pm 0.17\pm 0.03$ and 
$\cpipi = -0.09\pm 0.15\pm 0.04$, which are consistent with our previous 
measurements.  These results do not confirm the observation of large 
\CP\ violation reported in Ref.~\cite{BelleSin2alpha2004}.

We are grateful for the excellent luminosity and machine conditions
provided by our \pep2\ colleagues, 
and for the substantial dedicated effort from
the computing organizations that support \babar.
The collaborating institutions wish to thank 
SLAC for its support and kind hospitality. 
This work is supported by
DOE
and NSF (USA),
NSERC (Canada),
IHEP (China),
CEA and
CNRS-IN2P3
(France),
BMBF and DFG
(Germany),
INFN (Italy),
FOM (The Netherlands),
NFR (Norway),
MIST (Russia), and
PPARC (United Kingdom). 
Individuals have received support from CONACyT (Mexico), A.~P.~Sloan Foundation, 
Research Corporation,
and Alexander von Humboldt Foundation.

\end{document}